\documentclass[10pt]{article}

\usepackage{amsmath}
\usepackage{amssymb}
\usepackage{graphicx}

\usepackage{cite}

\usepackage{color} 

\usepackage[labelfont=bf,labelsep=period,justification=raggedright]{caption}

\makeatletter
\renewcommand{\@biblabel}[1]{\quad#1.}
\makeatother

\date{}

\def \>{\rangle} 
\def \<{\langle} 

\begin{document}

\begin{flushleft}
{\Large
\textbf{Identifying Keystone Species in the Human Gut Microbiome from Metagenomic Timeseries using Sparse Linear Regression }
}
\\
Charles K. Fisher$^{1}$, 
Pankaj Mehta$^{1,\ast}$
\\
\bf{1} Department of Physics, Boston University, Boston, MA, United States
\\
$\ast$ E-mail: pankajm@bu.edu
\end{flushleft}

\section*{Abstract}

Human associated microbial communities exert tremendous influence over human health and disease. With modern metagenomic sequencing methods it is now possible to follow the relative abundance of microbes in a community over time. These microbial communities exhibit rich ecological dynamics and an important goal of microbial ecology is to infer the ecological interactions between species directly from sequence data. Any algorithm for inferring ecological interactions must overcome three major obstacles: 1) a correlation between the abundances of two species does not imply that those species are interacting, 2) the sum constraint on the relative abundances obtained from metagenomic studies makes it difficult to infer the parameters in timeseries models, and 3) errors due to experimental uncertainty, or mis-assignment of sequencing reads into operational taxonomic units, bias inferences of species interactions due to a statistical problem called ``errors-in-variables''. Here we introduce an approach, Learning Interactions from MIcrobial Time Series (LIMITS), that overcomes these obstacles. LIMITS uses sparse linear regression with boostrap aggregation to infer a discrete-time Lotka-Volterra model for microbial dynamics.  We tested LIMITS on synthetic data and showed that it  could reliably infer the topology of the inter-species ecological interactions. We then used LIMITS to  characterize the species interactions in the gut microbiomes of two individuals and found that the interaction networks varied significantly between individuals. Furthermore, we found that the interaction networks of the two individuals are dominated by distinct  ``keystone species'', \emph{Bacteroides fragilis} and \emph{Bacteroided stercosis}, that have a disproportionate influence on the structure of the gut microbiome even though they are only found in moderate abundance. Based on our results, we hypothesize that the abundances of certain keystone species may be responsible for individuality in the human gut microbiome. 

\section*{Author Summary}

Numerous symbiotic microbial species live in the human body, making up what is called ``the human microbiome''.  Recent experiments have demonstrated the importance of the human microbiome to health and disease, but little is currently known about the structure and dynamics of these microbial populations. Modern DNA sequencing techniques provide a window into the ecology of these communities, but many obstacles stand in the way of making full use of these data for mapping out the interactions between microbial species that organize these communities in space and time. In this work, we demonstrate that it is possible to overcome each of these obstacles in order to find out who interacts with whom in the human microbiome. Applying our algorithm to study the interactions between the species residing in the guts of two individuals reveals that two ``keystone'' species have a disproportionate influence on the structure of the gut microbiome even though they are only found in moderate abundance. Based on our results, we hypothesize that the abundances of certain keystone species may be responsible for individuality in the human gut microbiome. 

\section*{Introduction}

Metagenomic sequencing technologies have revolutionized the study of the human-associated microbial consortia making up the human microbiome. Sequencing methods now allow researchers to estimate the relative abundance of the species in a community without having to culture individual species \cite{turnbaugh2007human, yatsunenko2012human, caporaso2011moving}. These studies have shown that microbial cells vastly outnumber human cells in the body, and that symbiotic microbial communities are important contributors to human health \cite{turnbaugh2007human}. For example, a recent study by Ridaura et al \cite{ridaura2013gut} demonstrated that transplants of gut microbial consortia are sufficient to induce obesity in previously lean mice or to promote weight loss in previously obese mice, suggesting an intriguing hypothesis that the composition of the gut microbiome may also contribute to obesity in humans. Many other studies have found significant links between the composition of human-associated microbial consortia and diseases including cancer and austim spectrum disorder \cite{moore1995intestinal, ahn2013human, turnbaugh2008core, hajishengallis2012keystone, hsiao2013microbiota, ridaura2013gut, wu2009human}. Despite the recent revelations highlighting the importance of the microbiome to human health, relatively little is known about the ecological structure and dynamics of these microbial communities. 

A microbial community consists of a vast number of species, all of which must compete for space and resources. In addition to competition, there are also many symbiotic interactions where certain species benefit from the presence of other microbial species. For example, a small molecule that is secreted by one species can be metabolized by another \cite{klitgord2010environments}. These species interactions provide a window with which to view the ecology of a microbial community, and allow one to make predictions about the effect of perturbations on a population \cite{faust2012microbial}. For example, removing a species that engages in mutualistic interactions may diminish the abundance of other species that depend on it for survival. Given their utility for understanding the ecology of a community, there is tremendous interest in developing techniques to infer interactions between species from metagenomic data \cite{faust2012microbial, stein2013ecological, friedman2012inferring}. 

There are two approaches to inferring dependencies between microbial species from metagenomic studies: cross-sectional analysis, and timeseries analysis \cite{faust2012microbial, stein2013ecological, friedman2012inferring}. Cross-sectional studies pool samples of the relative abundances of the microbial species in a particular environment (e.g.\ the gut) from multiple individuals and utilize correlations in the relative abundances as proxies for effective interactions between species.  By contrast, timeseries analysis follows the relative abundances of the microbial species in a particular environment, for a single individual, over time and utilizes dynamical modeling (e.g.\ ordinary differential equations) to understand dependencies between species.

Any methods for making reliable inferrences about species interactions from metagenomic studies must overcome three major obstacles. First, as shown below, a correlation between the abundances of two species does not imply that those species are interacting. Second, metagenomic methods measure the relative, not absolute, abundances of the microbial species in a community. This  makes it difficult to infer the parameters in timeseries models. Finally, errors due to experimental measurement errors and/or mis-assignment of sequencing reads into operational taxonomic units (OTUs), bias inferences of species interactions due to a statistical problem called ``errors-in-variables'' \cite{fuller1980properties}. We will show that each of these obstacles can be overcome using a new method we call LIMITS (Learning Interactions from MIcrobial Time Series). LIMITS obtains a reliable estimate for the topology of the directed species interaction network by employing sparse linear regression with bootstrap aggregation (``Bagging'') to learn the species interactions in a discrete-time Lotka-Volterra (dLV) model of population dynamics from a time series of relative species abundances \cite{breiman1996bagging, hofbauer1987coexistence}.


\section*{Results}

\subsection*{Correlation does not imply interaction}
Many previous works use the correlation between the relative abundances of two microbial species in an environment (e.g.\ the gut) as a proxy for how much the species interact.  In particular, a high degree of correlation between the abundances of two species is often taken as a proxy for a strong mutualistic interaction, and large anti-correlations, as indicative of a strong competitive interactions. Using correlations as a proxy for interactions suffers from several drawbacks. First, there are important subtleties involved in calculating correlations between species from relative abundances, but previous studies have presented algorithms (e.g.\ SparCC) to mitigate these problems \cite{friedman2012inferring}. More importantly, the abundances of two species may be correlated even if those species do not directly interact. For example, if species A directly interacts with species B, and species B directly interacts with species C, the abundances of species A and C are likely to be correlated even though do not directly interact. Finally, since correlation matrices are necessarily symmetric, all interactions learned using correlations also be symmetric.

\begin{figure}[!ht]
\begin{center}
\includegraphics[width=6in]{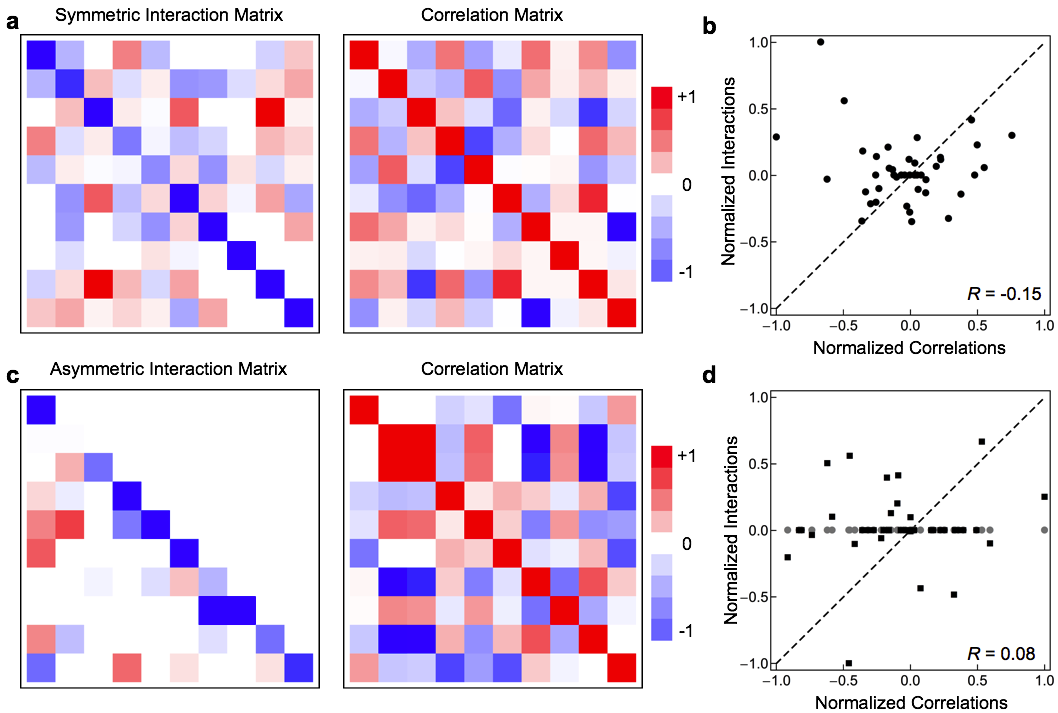}
\end{center}
\caption{ \textbf{ There is no simple relation between interaction coefficients and correlations in abundance. }  a) A symmetric interaction matrix and the corresponding correlation matrix. b) There is no relation between the interaction parameters and the correlations in abundance for the symmetric interaction matrix. c) An asymmetric interaction matrix and the corresponding correlation matrix. d) There is no relation between the interaction parameters and the correlations in abundance for the asymmetric interaction matrix. Points from above the diagonal in the interaction matrix are gray circles, whereas points from below the diagonal are black squares. In a and c, matrix elements have been scaled so that the smallest negative element is $-1$, the largest positive element is $+1$, and all elements retain their sign. In b and d, interaction coefficents were scaled so that the largest element by absolute value has $|c_{ij}| = 1$.}
\label{fig:fig1}
\end{figure}

The problems with using correlations in species abundances as proxies for species interactions can be illustrated with a simple numerical simulation. We used the dLV model (Eq.\ \ref{eq:dLV}) to simulate timeseries of the absolute abundances of 10 species for 1000 timesteps, starting from 100 different initial conditions, for two arbitrary species interaction matrices. Figure \ref{fig:fig1} compares the Pearson correlation matrices calculated from the absolute species abundances obtained from the dLV simulations and the true interaction matrices. It is clear from Fig.\ \ref{fig:fig1} that there is no obvious relationship between the correlations and the species interactions. That is, correlations in species abundances are actually very poor proxies for species interactions. 

In general, the  relationship between the interaction coefficients ($c_{ij}$) and the correlations in the species abundances is described by a complicated non-linear function that is difficult to compute or utilize. This can be seen by considering the linearized dynamics (Eq.\ \ref{eq:dLV}) of  $\ln x_i(t)$ around their equilibrium values, $\ln \< x_i(t) \>$. This stochastic process is a first-order autoregessive model described by 
\begin{equation}
\ln x(t +1) = \omega +  J \ln x(t) + \eta(t),
\end{equation}
where $\omega_i = -\sum_j c_{ij} \< x_j \> $, and $J_{ij} = \delta_{ij} + c_{ij} \<x_j\>$ is the Jacobian obtained by linearizing around the equilibrium species abundances. If $V$ is the covariance matrix with elements $V_{ij} = \< ( \ln x_i(t) - \< \ln x_i(t) \> ) ( \ln x_j(t) - \< \ln x_j(t) \> ) \>$ and $\Sigma$ is the covariance matrix of the Gaussian noise $\eta(t)$, then $\text{vec}(V) = (I_{n^2} - J \otimes J)^{-1} \text{vec}(\Sigma)$, where $\otimes$ is the Kronecker product and $\text{vec}$ is the matrix vectorization operator \cite{enders2008applied}. Thus, the interaction matrix is related to the covariance matrix by complex relation even in the linear regime of the dynamics. For this reason, it is very difficult, if not impossible, to determine the interaction coefficients using only knowledge of the correlations and equilibrium abundances.
 
Cross-sectional studies that pool data across individuals and utilize the correlations between the abundances of different species as proxies for species dependencies are especially affected by this problem. This suggests that time-series data is likely to be more suited for inferring ecological interactions than cross-sectional data. 

\subsection*{Timeseries inferrence with relative species abundances}

Even though the species interaction coefficients cannot be inferred from the correlations in species abundances, it is possible to reliably infer the interaction matrix using timeseries models. To do so, one utilizes a discrete time Lotka-Volterra Model (dLV) that relates the abundance of species $i$ at a time $t+1$ ($x_i(t+1)$) to the abundances of \emph{all} the species in the ecosystem at a time $t$ ($\vec{x} =\{x_1(t), \ldots, x_N(t) \}$). These interactions are encoded in the dLV through a set of interaction coefficients,  $c_{ij}$, that describe the influence species $j$ has on the abundance of species $i$ \cite{hofbauer1987coexistence}, and inferring these interaction coefficients is the major goal of this work.  The effect of species $j$ on species $i$ can be beneficial ($c_{ij} >0$), competitive ($c_{ij} < 0$), or the two species may not interact ($c_{ij} = 0$).

As shown in the Materials and Methods, given a time-series of the absolute abundances of the microbes in an ecosystem, one can learn the interaction coefficients by performing a  linear regression of $\ln x_i(t + 1) - \ln x_i(t)$ against $\vec{x}(t) - \< \vec{x} \>$, where $\<x\>$ is the vector of the equilibrium abundances.  It is important to note that each of these linear regressions can be performed independently for species $i = 1, \ldots, N$. In the following, we assume that the population dynamics are stable and that the equilibrium $\< x \>$ (or $\< \tilde{x} \>$) can be estimated by taking the median species abundances over the time series. 

Recall that most modern metagenomic techniques can only measure the relative abundances of microbes, not absolute abundances. This introduces additional complications into the problem of inferring species interactions using timeseries data. Although it is straight forward to infer species interactions by applying linear regression to a timeseries of absolute abundances, it is not \emph{a priori} clear that linear regression still works when applied to a timeseries of the relative abundances.  An important  technical problem that arises when using relative abundances is that the design matrix for the regression is singular because of the sum constraint on relative abundances of species ($\sum_i \tilde{x}_i(t) = 1$). As a result, there is no unique solution to the ordinary least squares problem applied to timeseries of relative species abundances. Nevertheless, the design matrix can be made to be invertible if one, or more, of the species are not included as variables in the regression. 

\begin{figure}[!ht]
\begin{center}
\includegraphics[width=6in]{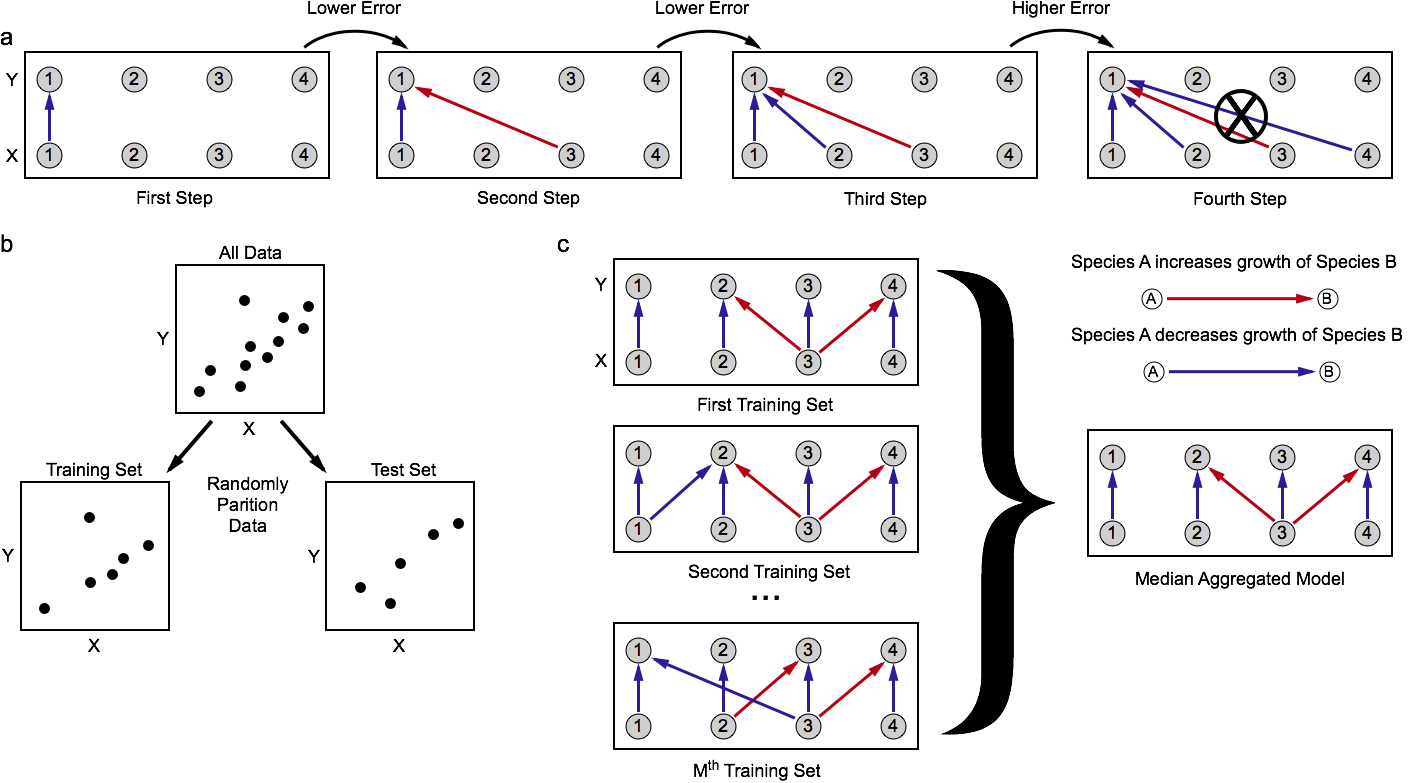}
\end{center}
\caption{ \textbf{Schematic illustrating forward stepwise regression and median bootstrap aggregating. } a) In forward stepwise regression, interactions are added to the model one at a time as long as including the additional covariate lowers the prediction error by a pre-defined threshold. b) The prediction error used for variable selection is evaluated by randomly partitioning the data into a training set used for the regression and a test used to evaluate the prediction error. c) Multiple models are built by repeatedly applying forward stepwise regression to random partitions of the data, each containing half the data points. The models are aggregated, or ``bagged'', by taking the median, which improves the stability of the fit while preserving the sparsity of the inferred interactions.   }
\label{fig:fig2}
\end{figure}

This insight motivates the use of a forward stepwise regression for selecting the covariate species that explain the changes in abundance of species $i$. In such a procedure, interactions and species are added sequentially to the regression as long as they improve the predictive power of model (see Fig.\ \ref{fig:fig2}a). Since the design matrix now only contains a sub-set of all possible species, it is never singular and the linear regression problem is well-defined. Furthermore, the goal of forward selection is to include only the strongest, most important species interactions in the model. Therefore, the resulting interaction networks are sparse and, hence, easily interpretable.

The procedure for forward stepwise regression is illustrated in Fig.\ \ref{fig:fig2}a. We know that each species must interact with itself, so $c_{ii}$ is the only interaction coefficient allowed to be nonzero in the first iteration. In each subsequent iteration, one additional interaction $c_{ij}$ is included in the model by scanning over all other species and choosing the one that produces the lowest error at predicting a test dataset. This is repeated as long as the prediction error decreases by a pre-specifed percentage that controls the sparsity of the model. A larger prediction error threshold results in more sparse solutions. 

Forward stepwise regression is a greedy algorithm, which results in a well-known instability \cite{breiman1996bagging}. This instability can be `cured' using a method called bootstrap aggregation, or ``Bagging'' (Fig.\ \ref{fig:fig2}b,c) \cite{breiman1996bagging}. To bag forward stepwise regression, the data are randomly partitioned into a training set used for the regression and a test set used for evaluating the prediction error. The prediction error threshold is a percentage that refers to how much the mean squared error evaluated on the test dataset must decrease in order to include an additional variable in the regression. The random partitioning of the data into training and test sets is repeated many times, each one resulting in a different estimate for the interaction matrix. The classical approach to Bagging calls for averaging these different estimates, but this destroys the sparsity of the solution. For this reason,  we use the median of the estimates instead of averages. This still greatly improves the stability of the inferred interaction matrix but preserves its sparsity. We call our algorithm LIMITS  (Learning Interactions from MIcrobial Time Series). 

\begin{figure}[!ht]
\begin{center}
\includegraphics[width=6in]{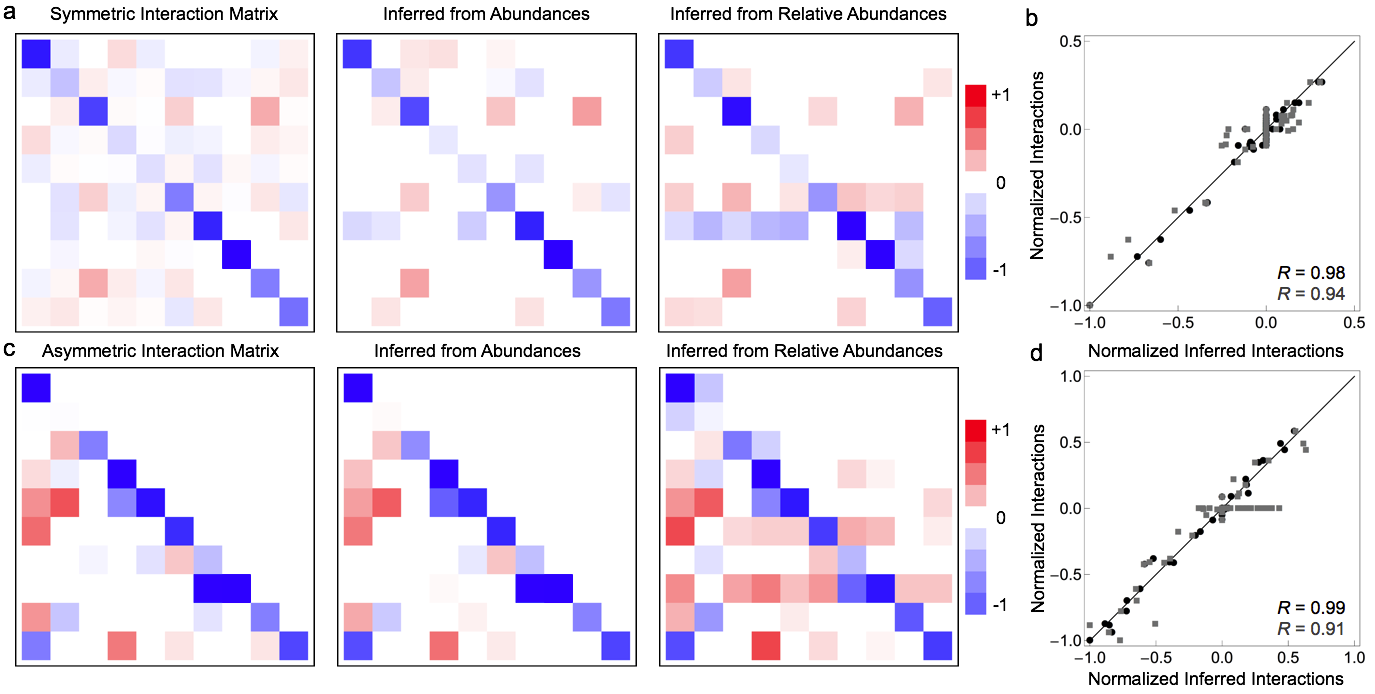}
\end{center}
\caption{ \textbf{ Example fits of interaction parameters using sparse linear regression. }  a) A symmetric interaction matrix (left), the corresponding matrix inferred from absolute abundance data (middle), and the corresponding matrix inferred from relative abundance data (right). b) There is good aggreement between the true and inferred interactions, from both absolute (black) and relative (gray) abundances, for the symmetric interaction matrix.  c) An asymmetric interaction matrix (left), the corresponding matrix inferred from absolute abundance data (middle), and the corresponding matrix inferred from relative abundance data (right). d) There is good aggreement between the true and inferred interactions, from both absolute (black) and relative (gray) abundances, for the asymmetric interaction matrix. The prediction error threshold was set to 5\% in for all fits. }
\label{fig:fig3}
\end{figure}

Figure \ref{fig:fig3} presents the results from applying LIMITS to infer the same interaction matrices discussed in Figure \ref{fig:fig1}. The data consist of either absolute or relative abundances from timeseries with 500 timesteps and 10 different initial conditions. The sample sizes are quite large, but not as large as used for the calculation of the correlations. The inferred parameters match the true interaction coefficients very accurately -- the smallest $R^2$ between the inferred and true parameters is 0.82 -- for both the symmetric (Fig.\ \ref{fig:fig3}a-b) and asymmetric (Fig.\ \ref{fig:fig3}c-d) interaction matrices using either absolute or relative species abundances.

\begin{figure}[!ht]
\begin{center}
\includegraphics[width=4in]{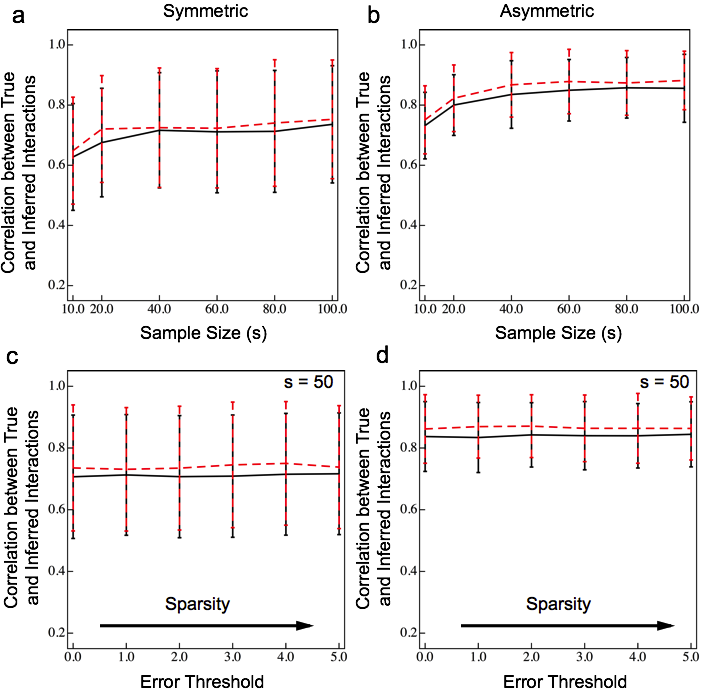}
\end{center}
\caption{ \textbf{ Performance of sparse linear regression as a functon of sample size and the prediction error threshold. }  a) Performance on absolute (red) and relative (black) abundances as a function of sample size for symmetric interaction matrices. b) Performance on absolute (red) and relative (black) abundances as a function of sample size for asymmetric interaction matrices. c) Performance on absolute (red) and relative (black) abundances as a function of the out-of-bag error threshold for symmetric interaction matrices. d) Performance on absolute (red) and relative (black) abundances as a function of the out-of-bag error threshold for symmetric interaction matrices. Error bars correspond to $\pm$ one standard deviation, and lines connect the means. }
\label{fig:fig4}
\end{figure}

To ensure that the exceptional performance of our sparse linear regression approach to inferring species interactions was not a fluke due to a particular choice of interaction matrices, we calculated the correlation between the true and inferred parameters for many randomly generated interaction matrices (Fig.\ \ref{fig:fig4}). Note that these simulations do not include any measurement errors (see next section for more on this). The performance of the algorithm obviously depends on sample size, but LIMITS generally perform admirably at inferring ecological interactions (see Figs.\ \ref{fig:fig4}a-b for symmetric and asymmetric matrices, respectively). Furthermore, Figs.\ \ref{fig:fig4}c-d show that our results do \emph{not} strongly depend on the prediction error threshold over most reasonable choices of threshold (0-5\%). These results demonstrate that in the absence of measurement noise,  LIMITS can successfully learn the interaction parameters from both absolute and  relative abundances.

\subsection*{Inferring interactions in the presence of measurement errors}

Up to this point, our analyses have ignored the impact of ``measurement noise'' on the inferred species interactions. There are two important sources of measurement noise in  metagenomic data. The first source is experimental noise introduced by sequencing errors. The second, and perhaps larger source of noise, is the mis-classification of sequencing reads into operational taxonomic units (OTUs). Most metagenomic studies rely on the sequencing of 16S rRNA to estimate species composition and diversity in a community. These 16S sequences are binned into groups, or OTUs,  that contain sequences with a predetermined degree of similarity. By comparing the sequences in an OTU to known sequences in an annotated database, it is often possible to assign OTUs to particular species or strains. In general, this is an extremely difficult bioinformatics problem  \cite{wooley2010primer} and is likely to be a significant source of measurement errors. Thus, any algorithm for inferring species interactions must be robust to measurement errors. 

At first glance, it  is tempting to assume that measurement noise, which we assume is multiplicative, simply adds to the stochastic ($\ln \eta_i(t)$) term that acts on the dependent variable ($\ln x_i(t+1) - \ln x_i(t)$) and, therefore, should have little impact on the inferred interactions (see Methods). However, as we discuss below, this is not the case since the $x_i(t)$ also act as the independent variables in the regression. Standard regression techniques assume that the independent variables are known exactly, and violation of this assumption results in biased parameter estimates even for asymptotically large sample sizes \cite{fuller1980properties}. For example, in the simplest case of a 1-dimensional regression $Y = \alpha + \beta X $ the estimator $\hat{\beta}$ is always less than the real $\beta$, i.e.\ $\hat{\beta} = \text{COV}(X,Y) / \text{VAR}(X) \leq \beta$ with equality only if there is no measurement error on $X$. The bias induced by using noisy indepedent variables in regression is known as the ``errors-in-variables'' problem in the statistical literature  \cite{fuller1980properties}. Analysis of the errors-in-variables bias for multivariate regression is more complicated than the 1-dimensional example; nevertheless, it can be stated quite generally that the interaction parameters inferred in the presence of signficant measurement errors will be incorrect. The most reliable method for mitigating the errors-in-variables bias is to measure additional data on some ``instrumental variables'' that provide information on the true values of the relative species abundances  \cite{fuller1980properties}. Unfortunately, in most cases we often do not have access to such additional data.

\begin{figure}[!ht]
\begin{center}
\includegraphics[width=4in]{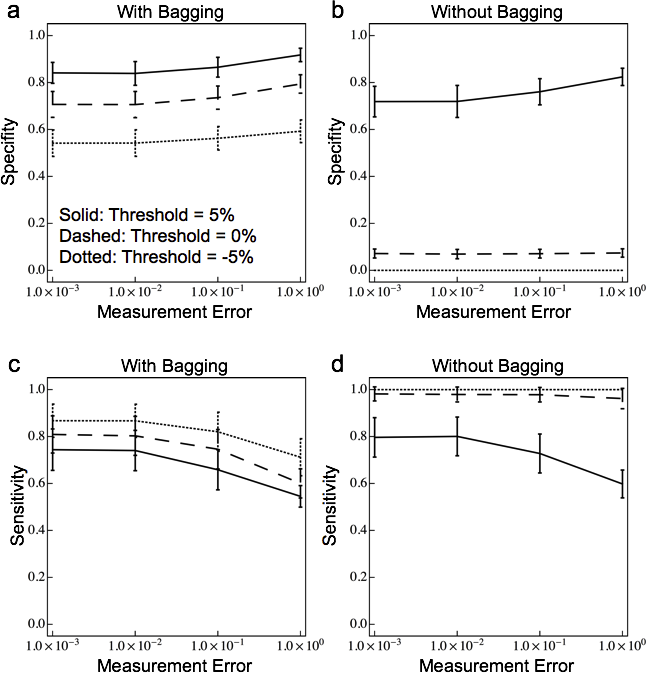}
\end{center}
\caption{ \textbf{ Sensitivity and specificity of predicted interactions as a function of measurement error for bagged and unbagged models. }  Specificity refers to the fraction of species pairs correctly identified as non-interacting, while sensitivity refers to the fraction of species pairs correctly identified as interaction. Both measures range from $0$ (poor performance) to $1$ (good performance). a) Specifity of sparse linear regression with Bagging as a function of measurement error for different prediction error thresholds. b) Specificity of sparse linear regression trained on the entire data set without Bagging as a function of measurement error for different prediction thresholds. c) Sensitivity of sparse linear regression with Bagging as a function of measurement error for different prediction error thresholds. d) Sensitivity of sparse linear regression trained on the entire data set without Bagging as a function of measurement error for different prediction error thresholds. Notice that without bagging, model performance is extremely sensitive to choice of prediction error threshold for adding new interactions }
\label{fig:fig5}
\end{figure}

Although the errors-in-variables bias cannot be eliminated, the topology of the interaction network can still be reliably inferred using our sparse linear regression approach even when the measurements of the relative species abundances are very noisy. Knowledge of which interactions are beneficial ($c_{ij} >0$), competitive ($c_{ij} < 0$), or zero ($c_{ij} = 0$) defines the topology of the interaction network. In the following simulations, we focus only on whether a given interaction is zero ($c_{ij} = 0$) or not ($c_{ij} \neq 0$) because we found that errors in the signs of the interactions were rare. The accuracy of the interaction topologies inferred from noisy relative abundance data were assessed using simulations with randomly chosen (asymmetric) interaction matrices. We computed the specificity -- the fraction of species pairs correctly identified as non-interacting -- and the sensitivity -- the fraction of species pairs correctly identified as interacting -- of the inferred topologies  (Fig.\ \ref{fig:fig5}) \cite{bishop2006pattern}. Figure \ref{fig:fig5}a shows that LIMITS produced specificities between 60\% and 80\% for different choices of the prediction error threshold, and that the performance was relatively insenstive to multiplicative measurement noise up to, and even beyond, 10\%. By contrast, Fig.\ \ref{fig:fig5}b shows that applying forward variable selection to the entire dataset, i.e.\ without Bagging, produced results that were very senstive to the choice of the prediction error threshold and even produced specificities as low as 0\%. The sensitivities for detecting interacting species with (Fig.\ \ref{fig:fig5}c) or without (Fig.\ \ref{fig:fig5}d) Bagging are both quite good, ranging between 70\% and 80\% for measurement noise up to 10\%. These results demonstrate that both the forward stepwise regression and the median bootrap aggregation are crucial components of the LIMITS algorithm. Moreover, LIMITS reliably infers the topology of the species interaction network even when there are significant errors-in-variables. 

\subsection*{Keystone species in the human gut microbiome}

\begin{figure}[!ht]
\begin{center}
\includegraphics[width=6in]{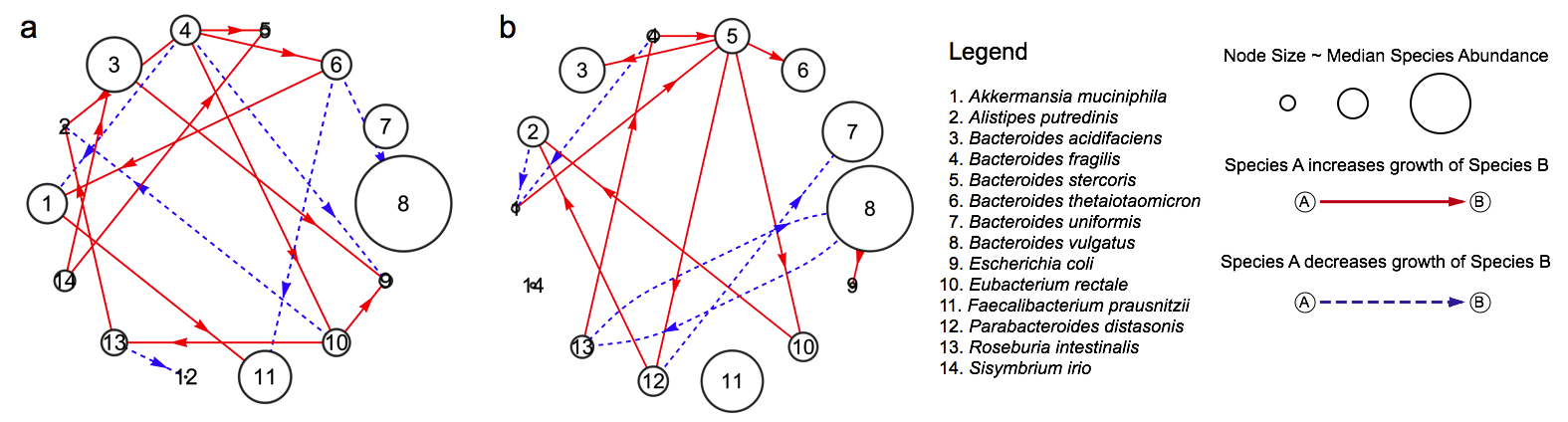}
\end{center}
\caption{ \textbf{ Interaction topologies of abundant species in the guts of two individuals. } The size of a node denotes the median relative species abundance, beneficial interactions are shown as solid red arrows, and competetive interactions are shown as dashed blue arrows. In individual a) species 4 \emph{Bacteroides fragilis} acts as a keystone species with 6 outgoing interactions, compared to a median number of outgoing interactions of $1$.  In individual b) species 5 \emph{Bacteroides stercosis} acts as a keystone species with 4 outgoing interactions, compared a median number of outgoing interactions of $1$. The 14 species included in the model were obtained by taking the union of the top 10 most abundant species from individuals a and b. The prediction error threshold was set to 3\%, graphs obtained using other prediction thresholds are shown in the Supporting Information. }
\label{fig:fig6}
\end{figure}

Emboldened by the success of our algorithm on synthetic data,  we applied LIMITS to infer the species interactions in the gut microbiomes of two individuals. The data from Caporaso et al \cite{caporaso2011moving} were obtained from the MGRAST database \cite{glass2010using} in a pre-processed form; i.e.\ as relative species abundances instead of raw sequencing data. These data consisted of approximately a half-year of daily sampled relative species abundances for individual (a) and a full year of daily sampled relative species abundances for individual (b). In general, we require at least $n^2$ timepoints in order to infer the interactions from $n$ species. The number of available timepoints was $O(100)$, so we considered only the $10$ most abundant species from individuals (a) and (b). Because the most abundant species were not entirely the same in individuals (a) and (b), we studied $14$ species obtained by taking the union of the $10$ most abundant species from individual (a) with the $10$ most abundant species from individual (b). The 14 species are listed in Fig.\ \ref{fig:fig6}. 

The species interaction network of the gut microbiome of individual (a) (shown in Fig.\ \ref{fig:fig6}a) is dominated by the species \emph{Bacteroides fragilis} even though it is found only in moderate abundances. \emph{Bacteroides fragilis} has 6 outgoing interactions in individual (a), in contrast to the other 13 species that have 0-3 outgoing interactions. The species interaction network of the gut microbiome of individual (b) (shown in Fig.\ \ref{fig:fig6}b) is also dominated by a single species, \emph{Bacteroides stercosis}, which is also found only in moderate abundaces. \emph{Bacteroides stercosis} has 4 outgoing interactions in individual (b), in contrast to the other 13 species that have 0-2 outgoing interactions. In addition, many of the interactions involving \emph{Bacteroides fragilis} and \emph{Bacteroides stercosis} are beneficial interactions. Based on these results, we refer to \emph{Bacteroides fragilis} and \emph{Bacteroides stercosis} as ``keystone species'' of the human gut microbiome because these two species exert tremendous influence on the structure of the microbial communities, even though they have lower median abundances than some other species. 

Additionally, we observed that the species interaction topology of the gut microbiome of individual (a) differs substaintially from the species interaction topology of the gut microbiome of individual (b), as is clear from Fig.\ \ref{fig:fig6}. In individual (a), \emph{Bacteroides fragilis} is much more abundant than \emph{Bacteroides stercosis} and, in turn, it is \emph{Bacteroides fragilis} that dominates the interaction network of individual (a). Likewise, in individual (b), \emph{Bacteroides stercosis} is much more abundant than \emph{Bacteroides fragilis} and, in turn, it is \emph{Bacteroides stercosis} that dominates the interaction network of individual (b). This observation motivates us to propose an intriguing hypothesis, that the abundances of certain keystone species are responsible for individuality of the human gut microbiome. Of course, much more data, from a larger population, will be required to confirm or reject this hypothesis. 

\section*{Discussion}

Metagenomic methods are providing an unprecedented window into the composition and structure of micriobial communities. They are revolutionizing our knowledge of microbial ecology and highlight the important roles played by the human microbiome in health and disease. Nevertheless, it is important to carefully consider the tools used to analyze these data and to address their associated challenges. We have highlighted three major obstacles that must be addressed by any study designed to use metagenomic data to analyze species interactions: 1) a correlation between the abundances of two species does not imply that those species are interacting, 2) the sum constraint on the relative abundances obtained from metagenomic studies makes it difficult to infer the parameters in timeseries models, and 3) errors due to experimental uncertainty, or mis-assignment of sequencing reads into operational taxonomic units (OTUs), bias inferrences of species interactions due to a statistical problem called ``errors-in-variables''. 

To overcome these obstacles, we have introduced a novel algorithm, LIMITS, for inferring species interaction coefficients that combines sparse linear regression with bootstrap aggregation (Bagging). Our method provides reliable estimates for the topology of the species interaction network even when faced with significant measurement noise. The interaction networks constructed using our approach are sparse, including only the strongest ecological  interactions. Regularizing the inference of the interaction network by favoring sparse solutions has the benefit that the results are easily interpretable, enabling the identification of keystone species with many important interactions. Furthermore, our  work suggests that it is difficult to learn species interactions from   cross-sectional studies that pool samples of the relative abundances of the microbial species from multiple individuals. This highlights the importance of collecting extended time-series data for understanding microbial ecological dynamics. 

We applied LIMITS to time-series data to infer ecological interaction networks of two individuals and found that the interaction networks are dominated by distinct keystone species. This motivated us to propose a hypothesis: that the abundances of certain keystone species are responsible for individuality of the human gut microbiome. While more data will be required to confirm or reject this hypothesis, it is intriguing to examine its potential consequences for the human microbiome. The keystone species hypothesis implies that even small perturbations to an environment can have a large impact on the composition of its resident microbial consortia if those perturbations affect a small number of important ``keystone'' species. Moreover, relatively small differences in individual diets, or minor differences in the interaction between the host immune system and the gut microbiota, that affect keystone species may be sufficient to organize gut microbial consortia into distinct types of communities, or ``enterotypes'' \cite{arumugam2011enterotypes, turnbaugh2009effect}. 

Our analysis identified the closely related species \emph{Bacteroides fragilis} and \emph{Bacteroides stercosis} as potential keystone species of the gut microbiome \cite{shah1989proposal}. Previous studies have suggested that the abundance of \emph{Bacteroides fragilis} modulates the levels of several metabolites and, in turn, the composition of the gut microbiome in a mouse model of gastrointestinal abnormalities associated with autism spectrum disorder \cite{hsiao2013microbiota}. Abundance of both \emph{Bacteroides fragilis} and \emph{Bacteroides stercosis} are associated with an increased risk of colon cancer \cite{moore1995intestinal, ahn2013human, wu2009human}, and previous authors even suggest that \emph{Bacteroides fragilis} acts as a critical ``keystone pathogen'' in the development of the disease \cite{hajishengallis2012keystone}. Classical ecological models of species interaction demonstrate that the manner of the interaction between two species is not solely a function of their identity, but is highly dependent on the environment in which the interaction takes place \cite{macarthur1970species, chesson2008interaction}.  Increased abundance of \emph{Bacteroides} species is associated with high fat diets, including typical Western diets with a high consumption of red meat that are associated with increased cancer risk \cite{moore1995intestinal}. It is possible that \emph{Bacteroides fragilis} and \emph{Bacteroides stercosis} act as keystone species in individuals consuming high fat diets due to their ability to convert bile into metabolites that are used by other members of the microbial community \cite{moore1995intestinal, hsiao2013microbiota}. 

The keystone species hypothesis can be experimentally tested by perturbing the abundance of individual species in a microbial consortium and observing the effect on the composition of the community. Our prediction is that most perturbations will have little impact on the overall structure of the microbial community, but perturbations applied to a small number of keystone species will have a large impact on the structure of the community. Due to ethical concerns, it is difficult to envision a direct experimental test of the keystone species hypothesis in human microbiota and, therefore, to test our specific predictions in regards to the keystone species \emph{Bacteroides fragils} and \emph{Bacteroides stercosis}. Nevertheless, experimental tests could be performed in animal models, or even in culture if a large enough microbial consortia can be assembled.

\section*{Materials and Methods}
\subsection*{Discrete Time Lotka-Volterra Model for absolute abundances}

Metagenomic sequencing methods have made it possible to follow the time evolution of a microbial population by determining the relative abundances of the species in a community in discrete intervals (e.g.\ one day). Given the discrete nature of these data, it is most sensible to use a discrete-time model of population dynamics. The discrete-time Lotka-Volterra (dLV) model of population dynamics (sometimes called the Ricker model) relates the abundance of species $i$ at a time $t+\delta t$ ($x_i(t+\delta t)$) to the abundances of \emph{all} the  species in the ecosystem at a time $t$ ($\vec{x}= \{x_1(t) \ldots x_N(t) \}$). These interactions are encoded in the dLV through a set of interaction coefficients,  $c_{ij}$,  that describe the influence species $j$ has on the abundance of species $i$ \cite{hofbauer1987coexistence}, and inferring these interaction coefficients is the major goal of this work. The abundance of a species can also change due to environmental and demographic stochastic effects. The dLV can be generalized to include stochasticity by including  a log-normally distributed multiplicative noise, $\eta_i(t)$.  Specifically, the dynamics are modeled by the equations 
\begin{equation}
\label{eq:dLV}
x_i(t + \delta t) = \eta_i(t) x_i(t) \exp( \delta t \sum_j c_{ij} (x_j(t) - \<x_j\>) ), 
\end{equation}
where $\<x_j\>$ is equilibrium abundance of species $j$ and is set by the carrying capacity of the environment. In writing these equations in this form, we have assumed that in the absence of noise the dLV equations have a unique steady-state solution with the abundances given by $\<x_j\>$.
Notice that in the absence of multiplicative noise and the limit $\delta t \rightarrow 0$, the dLV model reduces  to the usual  Lotka-Volterra differential equations: 
\begin{equation}
\frac{1}{x_i(t)} {d x_i(t) \over dt} =  \sum_j c_{ij}(x_j(t) - \<x_j\>).
\end{equation}

In what follows, without loss of generality, we set $\delta t=1$. This is equivalent to measuring time in units of $\delta t$.  To fit microbial data, it is actually helpful to work with the logarithm of Equation (\ref{eq:dLV}). Furthermore, we assume that the sampling time and the update time are both equal to 1. Taking the logarithm  of Eq.\ \ref{eq:dLV} yields,
\begin{equation}
\ln{x_i(t +1)} -\ln{x_i(t)} = \zeta_i(t) +\sum_{j} c_{ij} (x_j(t) - \<x_j\>),
\end{equation}
where by construction $\zeta_i(t) = \ln{\eta_i(t)}$ is a normally distributed variable. This logarithmic form of the dLV model is especially convenient for inferring species interactions from time series of species abundances because the inference problem reduces to standard linear regression (as discussed below).

Thus, far we have assumed that it is possible to directly measure the absolute abundances $x_i(t)$. However, in practice, metagenomic sequencing studies typically provide relative abundances $\tilde{x}_i = Z^{-1} x_i$ , where $Z = \sum_j x_j$. Provided that the number of species is large ($N \gg 1$), the fluctuations in the total population size $Z(t) = \sum_i x_i(t)$ around its mean value $\<Z\> = \sum_i \<x_i\>$  will be small. In this case, the dynamics of the relative abundances ($\tilde{x}_i(t)$) are well-described by a modified dLV model:
\begin{equation}
\label{eq:dLV2}
\tilde{x}_i(t + 1) \approx \eta_i(t) \tilde{x}_i(t) \exp( \sum_j \tilde{c}_{ij} (\tilde{x}_j(t) - \<\tilde{x}_j\>) ),
\end{equation}
where we have defined the new interaction coefficients  $\tilde{c}_{ij} = \< Z \> c_{ij}$ which are related to the true interaction coefficients $c_{ij}$ by the average population size. Thus, relative species abundance data can be modeled using the dLV model, but the interaction coefficients are known only up to an arbitrary multiplicative constant. However, as discussed in the main text, the design matrix for relative species abundances is singular so simple linear regression fails. 

In all of the simulations discussed in the main text the stochasticity was set to $\ln \eta_i(t) \sim \mathcal{N}(0,0.1)$.

\subsection*{Linear Regression}

Suppose we are given data consisting of the absolute (or relative) abundances of the species in a population of $N$ species in the form of timeseries of length $T$ starting from $M$ initial conditions. We infer each row ($\vec{c}_i = \{c_{ij}\}_{j=1}^N$) of the interaction coefficient matrix ($C$) separately. The equilibrium population $\< x \>$ (or $\< \tilde{x} \>$) is assumed to correspond to the population median. The design matrix is an $M (T-1) \times N$ matrix with rows $X_l = \{ x_1^{(k)} (t) - \<x_1\>, ... , x_N^{(k)}(t) - \< x_N \>\} $. The data vector has length $M (T-1)$ and is given by $\vec{v}_i = \{ \ln x_i^{(k)}(1) - \ln x_i^{(k)}(0) , ... , \ln x_i^{(k)}(T) -\ln x_i^{(k)}(T-1) \}$. Note that any of the timepoints with $x_i^{(k)}(t) = 0$ were left out of the regression because the logarithm of zero is undefined. The least squares estimate for the interaction coefficients is $\hat{c}_i = X^+ \vec{v}_i$, where the $+$ denotes the psuedo-inverse.

\subsection*{Outline of the LIMITS Algorithm}

Here, we present a high-level outline of the LIMITS algorithm (see Fig.\ \ref{fig:fig2}). An implementation of the LIMITS algorithm written in Mathematica (Wolfram Research, Inc.) is available from the authors upon request. Since all of the regressions are performed independently for each species, we will only describe the algorithm for inferring a row of the interaction matrix ($\vec{c}_i = \{c_{ij}\}_{j=1}^N$). One simply loops over $i$ to obtain the full interaction matrix. Moreover, bootstrap aggregating simply involves performing the whole proceedure $L$ times, thereby constructing multiple estimates $\vec{c}_i^{(1)}, \ldots, \vec{c}_i^{(L)}$ and taking their median. Thus, the only thing that takes some effort to explain is how to construct one of estimate of $\vec{c}_i$.

\begin{enumerate}
\item First, we randomly partition the data into a training set and a test set, each containing half the data points. 
\item A set of active coefficients is initialized to $\text{ACTIVE} = \{ c_{ii} \}$ and a set of inactive coefficients is initialized to $\text{INACTIVE} = \{ c_{ij} \}_{j \neq i}$. 
\item A linear regression including only species $i$ is performed on the training set, and the inferred coefficient is used to calculate a prediction error (called $\text{ERROR}$) for the test dataset. 
\item For each coefficient $c$ in $\text{INACTIVE}$ create $\text{TEST} = \text{ACTIVE} \bigcup \{c\}$ and perform a linear regression using the coefficients in $\text{TEST}$ against the training dataset. 
\item Next, the inferred coefficients are used to calculate the prediction errors for the test dataset. The particular species $j$ with the smallest prediction error is retained, and we call this error $\text{ERROR}(j)$. 
\item If $100 \times ( \text{ERROR}- \text{ERROR}(j)) / \text{ERROR}$ is greater than a pre-specified error threshold then we set $\text{ERROR} = \text{ERROR}(j)$, $\text{ACTIVE} = \text{ACTIVE} \bigcup \{c_{ij}\}$, and $c_{ij}$ is deleted from $\text{INACTIVE}$, otherwise we terminate the loop and return an estimate for the interactions $\vec{c}_i= \{c_{ij}\}_{j=1}^N$.
\end{enumerate}

\section*{Acknowledgments}
We would like to acknowledge useful conversations with Sara Collins and Daniel Segr\'{e} and thank Alex Lang and Javad Noorbakhsh for comments on the mansuscript. 
PM and CKF were partially supported by a Sloan Research Fellowship.

\bibliography{keystone_refs}

\section*{Figure Legends}

\setcounter{figure}{0} \renewcommand{\thefigure}{S\arabic{figure}} 

\begin{figure}[!ht]
\begin{center}
\includegraphics[width=6in]{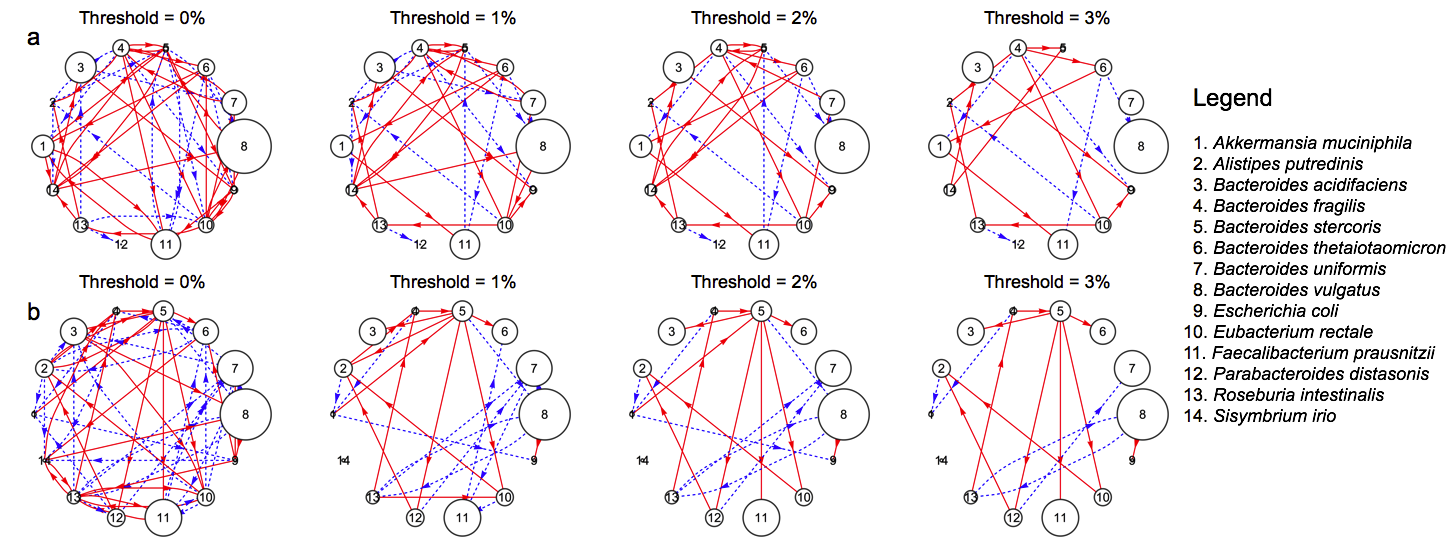}
\end{center}
\caption{ \textbf{Interaction topologies of abundant species in the guts of two individuals using different prediction error thresholds. } The size of a node denotes the median relative species abundance, beneficial interactions are shown as solid red arrows, and competetive interactions are shown as dashed blue arrows. The 14 species included in the model were obtained by taking the union of the top 10 most abundant species from individuals a and b. }
\label{fig:figS1}
\end{figure}



\end{document}